\documentclass[final,5p,times,twocolumn]{elsarticle}

\usepackage{graphicx}

\usepackage{amssymb}
\usepackage{amsthm}

\usepackage{lineno}

\journal{Physics Letters B}

\begin{document}

\begin{frontmatter}

\title{Analysis of clustering phenomena in ab initio approaches}

\author[label1,label2,label3]{D. M. Rodkin\corref{cor1}}
\address[label1]{Dukhov Research Institute for Automatics, 127055, Moscow, Russia}
\address[label2]{Moscow Institute of Physics and Technology, 141701, 9 Institutskiy per., Dolgoprudny, Moscow Region}
\address[label3]{Pacific National University, 680035, Khabarovsk, Russia}
\ead{rodkindm92@gmail.com}

\author[label1,label2,label3,label4]{Yu. M. Tchuvil'sky}
\address[label4]{Skobeltsyn Institute of Nuclear Physics, Lomonosov
Moscow State University, 119991 Moscow, Russia}
\ead{tchuvl@nucl-th.sinp.msu.ru}

\cortext[cor1]{I am corresponding author}

\begin{abstract}
An approach for explicit consideration of cluster effects in nuclear systems and
accurate {\it ab initio} calculations of cluster characteristics of nuclei is devised.
The essential block of the approach is a construction of a basis which incorporates both
conventional No-Core Shell Model wave functions and translationally-invariant wave
functions of various cluster channels. The results of computations of the total binding
energies of $^8$Be nucleus as well as the spectroscopic factors of cluster channels
(amount of clustering) are presented. An unexpected fresh result of the rigorous study
is that the contribution of "non-clustered" components of the basis to the total binding
energy is great even in the typical cluster systems such as the discussed nucleus.
\end{abstract}


\end{frontmatter}

One of the fundamental properties of light nuclei is the clustering displaying itself in
a certain degree of separation of a nucleus into two or more multi-nucleon
substructures. A great body of experimental information has been accumulated over many
years of studies of the clustering phenomena. Theoretical studies of clustering
originated in Ref. \cite{wheel}. A microscopic, i. e. starting from a certain
NN-potential and considering a nucleus or a two-fragment collision channel as an
A-nucleon or an (A$_1$ + A$_2$)-nucleon system -- so-called Resonating Group Model (RGM)
-- has been put forward in it. Forty-year evolution of these studies has been summarized
in Ref. \cite{wild}. A large number of microscopic approaches taking cluster properties
into account was discussed in the monograph. In the view of the authors 
one of the main lines
of nuclear theory is to construct a unified theory of nuclear structure and nuclear
collision dynamics in the framework of microscopic approaches.  The "dynamic" clustering
is peculiar to the collision processes therefore the nuclear reaction theory involve
cluster concepts almost without exceptions. In succeeding years a variety of theoretical
techniques have been developed to study nuclear clustering. Within modern microscopic
models such as the Generator Coordinate Method (GCM) \cite{hor,desc1}, Microscopic Cluster Model (MCM) \cite{desc2,desc3}, THSR-approach
\cite{thsr}, Antisymmetrized Molecular Dynamics (AMD) \cite{keh} and Fermionic Molecular
Dynamics (FMD) \cite{neff1,neff2}, clustering in various nuclear states has been confirmed to
emerge directly from NN-interactions. A detail discussion of these approaces is presented in review \cite{freer}. Large-scale calculations of cluster
characteristics of nuclear states: cluster spectroscopic factors and form factors have
been studied in the framework of advanced shell-model method -- so-called
Cluster-Nucleon Configuration Interaction Model \cite{avil,vt2,vt3,vt4}.

The supercomputing era provides new possibilities for building the unified theory. Due
to that the development of {\it ab initio} approaches to description of nuclear
structure and dynamics is recently one of the basic lines of the advancement of nuclear
science. Such approaches are based on Hamiltonians involving universal (common for wide
range of objects under study), realistic NN-, NNN-, etc. potentials. No-Core Shell Model
(NCSM) is one of the most advanced among these approaches. A typical basis of this model
contains all possible nucleon configurations on equal terms up to a certain truncation
level. Obviously in the most cases huge basis is required to achieve convergence of the
results. Recently various versions of NCSM \cite{nvb,nqsb,mar1,mar2,mar3,bar} occupy a
prominent place in nuclear structure calculations.

Microscopic calculations of light nuclei properties demonstrate "non-equivalence" of different nucleon configurations. For example many of Slater determinants play a negligible role in computations of 
nuclear total binding energy in NCSM. For this reason, methods of selection of essential
components of a certain nature, {\it a priori} or after preliminary analysis,  are recently popular \cite{lui,abe,dyt1,dr,tob}.

The studying of mentioned non-equivalence originated by clustering phenomena is, in our
opinion, an intriguing issue. In the current paper we carry out the analysis of the role
of cluster components in solutions of the A-nucleon Schr\"odinger equation with {\it ab
initio} NN-potentials. For these purposes we combine the wave functions (WFs) of various
cluster channels and standard NCSM components into unified basis. Another vital issue is
concerning the realistic numerical values of cluster characteristics -- spectroscopic
factors (SFs) of cluster channels.  These values carry information on "nuclear geometry"
as well as the information necessary for calculation of the reaction cross sections. In
nuclear structure studies SFs play a role of "amount of clustering", according to the
terminology of Ref. \cite{lov}. {\it Ab initio} approaches are required to compute these
values. In the framework of our studies we explore both these problems. We have
performed computations of the total binding energies (TBEs) of clustered nuclei as well
as the SFs of their fragmentation channels in the framework of different models in which
the clustering is taken into account in a number of ways. We also introduce a definition
of aggregate amount of clustering (AoC) useful in the case when the multi-channel
problem is considered. This mathematical object is non-trivial owing to the strong
non-orthogonality of different cluster channels.

To design the desired formalism we build up partial bases of translationally-invariant
A-nucleon WFs of channels manifesting two-fragment separation A = A$_1$ + A$_2$. Various
channels are distinguished by the internal state WFs of the fragments A$_1$ and A$_2$ as
well as the channel spin and relative motion angular momentum. These WFs are {\it ab initio} calculated. The next problem is to assemble the non-orthogonal partial bases corresponding to these channels into a
unified orthogonal basis and to add a number of eigenvectors obtained in ordinary NCSM
calculations (called "polarization terms" in Ref. \cite{wild}) to this assembled cluster
basis. For this purpose the WFs of channels are transformed to the shell-model form and
undergo the orthonormalization procedure together with polarization terms. In such a way we build the universal basis suitable to describe arbitrary
states taking one- or multi-channel clustering into account. This potentiality is
topical especially for the states manifesting pronounced cluster properties.

It should be noted that an approach aimed at an accurate {\it ab initio} description of
cluster reactions induced by light nuclei collisions has been developed in Refs.
\cite{qua,nav1}. This approach exploits the microscopic Hamiltonians together with RGM
approach (so accounting for the "dynamic" clustering). It was called NCSM/RGM. The
"polarization terms" were introduced into NCSM/RGM in Refs. \cite{baro1,baro2,lang,doh,ncsmc}.
The new model received the name No-Core Shell Model with Continuum (NCSMC).

By contrast in the current paper we concentrate on manifestation of cluster structure in
bound and resonance states. Another difference between the NCSMC and our approach is
technical one. The technique of so-called cluster coefficients (CCs), presented in Refs. \cite{st1,st2,tch1,nem}, is used in our work
for transformation of the cluster WFs to the superpositions of the Slater determinants. It
provides a general approach to work with a broad variety of rather heavily and excited fragments. Besides
that applying this formalism we obtain pure algebraic approach which seems to be of a
great convenience.

Our calculations are carried out with the use of Hamiltonians containing high-precision
modern NN-potentials JISP16 \cite{shir2} and Daejeon16 \cite{shir1}. The first one is constructed using the J-matrix inverse scattering method. The latter one is
built using the N3LO limitation of Chiral Effective Field Theory \cite{mach} softened by
similarity renormalization group(SRG) transformation. SRG transformation had been
proposed in Refs. \cite{bfp}. Both these potentials are well-tested in broad spectral
calculations of A$\leq$16 nuclei.

Notice that a NCSM approach was already used in computation of the $\alpha$-cluster SFs
in $^8$Be nucleus where SRG-softened N3LO-based NN-interaction was used. Results of
these studies were presented in Ref. \cite{kv}. We involve these values of SFs in our
analysis.

Let us consider a two-fragment system A$_1$ + A$_2$. The oscillator-basis terms of the
cluster channel $c_\kappa$ are built in the translationally-invariant form:
\begin{equation}
\Psi^{c_\kappa, TI} _{{A\,}, nl}  = \hat A\{\Psi^{\{k_1\}} _{A\,_1 } \Psi^{\{k_2\}} _{A\,_2
} \varphi _{nl} (\vec \rho )\} _{JM_JT} , \label{eq1}
\end{equation}
 where $A = A_1  + A_2,\; \hat A$ is the antisymmetrizer, $\Psi^{\{k_i\}}
_{A\,_i}$ is a translationally-invariant internal WF of the fragment labelled by a set
of quantum numbers $\{k_i\}$; $\varphi _{nl} (\vec \rho )$ is the WF of the relative
motion. A channel WF as a whole is labelled by the set of quantum numbers $c_\kappa$ which includes
$\{k_1\},\{k_2\},J,M_J,T$. The goal of the subsequent transformations is to
present function (\ref{eq1}) as a linear combination of the Slater determinants (SDs)
containing the one-nucleon WFs of the oscillator basis. The reason of it was mentioned above. For these purposes function
(\ref{eq1}) is presented in the form of the linear combination of the WFs with fixed magnetic quantum numbers $m$, $M_{c_{\kappa}}$ and multiplied by the function of the center of mass (CM) zero vibrations 
$\Phi _{000} (\vec R)$. 
Then the transformation of WFs caused by changing from $\vec
R,\vec \rho$ to  $\vec R_1 ,\vec R_2$ coordinates -- different-mass Talmi-Moshinsky
transformation defined in \cite{sm}

\begin{equation}
\begin{array}{rcl}
\Psi^{c_\kappa} _{{A\,}, nl} = \Phi _{000} (\vec R)\Psi^{c_\kappa, TI} _{{A\,}, nl} =  \\[\bigskipamount]
\sum\limits_{N_i ,L_i ,M_i,m,M_{c_{\kappa}}}
{\left\langle {{\begin{array}{*{20}c}
   {000}  \\
   {nlm}  \\
\end{array}  }}
 \mathrel{\left | {\vphantom {{\begin{array}{*{20}c}
   {000}  \\
   {nlm}  \\
\end{array}  } {\begin{array}{*{20}c}
   {N_1 ,L_1 ,M_1 }  \\
   {N_2 ,L_2 ,M_2 }  \\
\end{array}}}}
 \right. \kern-\nulldelimiterspace}
 {{\begin{array}{*{20}c}
   {N_1 ,L_1 ,M_1 }  \\
   {N_2 ,L_2 ,M_2 }  \\
\end{array}}} \right\rangle } \\[\bigskipamount]
\hat A\{ \Phi _{N_1 ,L_1 ,M_1 }^{A_1 } (\vec R_1 )  
\Psi^{\{k_1\}}_{A\,_1 } \Phi _{N_2 ,L_2 ,M_2 }^{A_2 } (\vec R_2 )\Psi^{\{k_2\}} _{A\,_2
} \}_{JM_J } - \label{eq3}
\end{array}
\end{equation}
 is performed.

The main procedure of the method is to transform internal WFs
corresponding to each fragment with none-zero center of mass vibrations into a superposition of SDs
\begin{equation}
\Phi _{N_i ,L_i ,M_i }^{A_i } (\vec R_i )\Psi^{\{k_i\}}_{A\,_i }  = \sum\limits_k
{X_{N_i ,L_i ,M_i }^{A_i (k)} \Psi _{A\,_i (k)}^{SD} }. \label{eq4}
\end{equation}
Quantity $X_{N_i ,L_i ,M_i }^{A_i (k)}$ is called a cluster coefficient.  There is a large number of methods
elaborated for the calculations of CCs (see Refs. \cite{st1,st2,tch1,nem} ). The most general scheme is based on the method
of the second quantization of the oscillator quanta. In this scheme the WF of the CM
motion is presented as

\begin{equation}
\Phi _{N_i ,L_i ,M_i }^{A_i } (\vec R_i ) = N_{N_i ,L_i } (\hat{ \vec \mu^{\dag}})^{N_i - L_i
} Y_{L_i,M_i} (\hat{ \vec \mu^{\dag}})\Phi_{000}^{A_i} (\vec R_i ), \label{eq5}
\end{equation}
where  $\hat{ \vec \mu^{\dag}}$ is the creation operator of the oscillator quantum, and
$N_{N_i,L_i }$ is the norm constant. Thus the CC turns out to be reduced to a matrix
element of the tensor operator $ \hat{ \vec \mu^{\dag}}$:

\begin{equation}
\begin{array}{rcl}
 < \Psi _{A_i (k)}^{SD} |\phi _{N_i ,L_i } (\vec R_{A_i } )\Psi^{\{k_i\}}_{A_i }  >  =
 N_{N_i ,L_i } 
  \left\langle {\Psi _{A\,_i (k)}^{SD} }
 \right|  \\
 (\hat \mu ^\dag  )^{N_i  - L_i } 
Y_{L_i,M_i}(\hat{ \vec \mu^{\dag}}) 
 \left| {\Phi _{000}^{A_i } (\vec R_i )\Psi^{\{k_i\}}_{A\,_i } } \right\rangle
\end{array}
\label{eq6}
\end{equation}

Contrary to pioneering work \cite{kurd} in which translationally-invariant WFs were
written in terms of Jacobi coordinates, the formula

\begin{equation}
\Psi^{c_\kappa, TI} _{{A\,}, nl} = \Psi^{c_\kappa} _{{A\,}, nl} /\Phi _{000}^{A} (\vec R)
\label{eq7}
\end{equation}
 is considered here as the definition of these functions.

It should be noted that in the general case WFs of cluster-channel terms (\ref{eq1}) of
one and the same channel $c_\kappa$ characterized by the pair of internal functions
$\Psi^{\{k_1\}}_{A_1}$, $\Psi^{\{k_2\}}_{A_2}$  and different values of relative motion
quantum numbers $n,l$ are non-orthogonal. The same is true for the terms of different
channels. Moreover all these WFs are non-orthogonal with the polarization terms - eigenvectors obtained in ordinary NCSM
calculations. So the
next step is to build a basis of orthonormalized WFs which includes both the
polarization terms and the terms of several cluster channels. The basis is obtained by
the diagonalization of matrix

\begin{equation}
\left\| {\begin{array}{*{20}c} \left[ {\left\langle \Psi_{pol}^{(j' )}  \right|\left.
\Psi _{pol}^{(j)} \right\rangle } \right] & \left[ {\left\langle \Psi_{pol}^{(j' )}  \right|\left. \Psi^{c_{\kappa'}}_{{A\,}, nl} \right\rangle } \right]  \\[\bigskipamount]
\left[ {\left\langle \Psi_{pol}^{(j' )}  \right|\left.
\Psi^{c_\kappa} _{{A\,}, nl} \right\rangle } \right] & \left[ {\left\langle \Psi^{c_\kappa} _{{A\,}, nl}  \right|\left. \Psi^{c_{\kappa'}} _{{A\,}, nl} \right\rangle } \right]  \\
\end{array}} \right\|
\label{eq8}
\end{equation}
in which the square brackets denote the sub-matrixes. The cluster-channel terms 
$\Psi^{c_\kappa} _{{A\,}, nl}$ are, in fact, expressed in the
form of superpositions of SDs with the help of formula (\ref{eq3}).

The eigenvectors of the matrix normalized by its eigenvalues give the desirable basis.
Each term of the basis takes the form of a SDs linear combination. Therefore the
computation of the matrix elements of both the kinetic and the potential energy in the
discussed basis is identical to the ordinary shell-model computation. An arbitrary
microscopic {\it ab initio} or effective, including two-, three-, etc. nucleon forces
Hamiltonian may be utilized. The calculations of the matrix elements of operators, the
estimates of error bars etc. are also analogous to those in the shell model. The only
difference is the list of the basis vectors. This list is considerably shorter compared
to the NCSM one. The limitations on the use of the approach are imposed by the
dimensionality of the basis vectors.

This approach as a whole is very adaptable due to the possibilities to vary: a number of
cluster channels and polarization terms; $n_{max}$; A-, A$_1$-, A$_2$-nucleon shell-model spaces
determined by the corresponding truncation level parameters ($N^{(i)}_{max}$) which are
the maximal values of the total number of the oscillator quanta in each subsystem. This
gives a way to take into account various halo, cluster and other properties of a system.

The formalism presented above is convenient for the calculation of the SFs of arbitrary 
solutions of A-nucleon Schr\"odinger equations $\Psi_A$. The SF of a certain cluster
channel $c_\kappa$, which may be called in the context of the current paper the
one-channel amount of clustering, is defined as the the sum of squared overlaps of the
wave function $\Psi_A$ with the normalized antisymmetric WFs $\Psi^{c_\kappa, orth}_{A(i)}$. WFs $\Psi^{c_\kappa, orth}_{A(i)}$ are  obtained by 
the diagonalization of sub-matrix contained in right-lower quadrant of the matrix
(\ref{eq8}) which is reduced by additional condition $\kappa=\kappa'$. The just
presented definition is completely equivalent to the one  proposed in Ref. \cite{flis1}
(so-called "new SF"). This definition plays an important role in the theory of nuclear
reactions. Detailed discussion of various aspects of this concept is presented in Refs.
\cite{lov,kkt,vt3,vt4}.

A treatment of the multi-channel problem is a more delicate problem. Obviously the
analysis of statistical weights of components contained in a WF is possible only with
the proviso that these components are orthogonal one to another. The basis of
cluster-channel terms (\ref{eq1}) incorporating all channels $c_\kappa$ of fragmentation
A$_1$ + A$_2$ (all internal states of these clusters) is complete and what is more
linear dependent. Orthonormalization of the basis mixes the terms of different channels and, because
of the linear dependency, the result of the procedure is ambiguous. Therefore a
possibility to estimate contributions of individual cluster channels correctly is
doubtful in principle. Nevertheless it is possible to perform the analysis of aggregate
AoC for chosen list of channels. For this purpose the definition of the AoC of a channel
can be generalized to the aggregate AoC of a group of channels. This value  is defined
similarly to the value of one-channel AoC but the sum of squared overlaps  of the
wave function $\Psi_A$ with the WFs
$\Psi^{c_\kappa, orth}_{A(i)}$ is over all terms of a chosen number of cluster channels
$\{c_\kappa\}$. WFs of the terms are obtained by the orthonormalization of the WFs $\Psi^{c_\kappa} _{{A\,}, nl}$ corresponding to all these channels. The definition allows one to
determine the measure of clustering depending upon the choice of the set of cluster
channels.

Notice that the choice of the maximal value of $n$ and A$_1$-, A$_2$-nucleon truncation
level parameters ($N^{(i)}_{max}$) may be different for calculation of the WFs of
nuclear states and for the evaluation of the SFs and the aggregate AoC.

Characteristics of several clustered systems were analyzed in our work. In this paper the
cluster properties of $^8$Be nucleus as $\alpha+\alpha$ system are presented to
demonstrate some unexpected aspects of clustering typical for all the systems. We
demonstrate here the values of the total binding energy (TBE), the one-channel and
aggregate AoC. Widely-used code Antoine rearranged by us for performing NCSM
computations is exploited to calculate the WFs of the clusters and the polarization
terms.

The TBEs of $^8$Be nucleus have been computed with the use of variously constructed
bases. In all instances the maximal total number of the oscillator quanta 
$N^{max}_{tot}$ is considered as a basic characteristic parameter together with
$\hbar\omega$ one.

 The bases are the following.
First, conventional basis of NCSM is used. Let us denote this version  as {\it mod1}
bellow. Second, two types of pure cluster bases are exploited. One of them contains the
WFs of the ground states of both $\alpha$-clusters $\Psi_{\alpha}$ in their lowest
shell-model configurations. The other one incorporates the realistic WFs of $0^+_1$ and
$0^+_2$ states of $^4$He calculated using the code Antoine with truncation level
$N^{max}_{\alpha}$=2. This basis is three-channel. These versions are denoted as {\it
mod2} and {\it mod3}. 
Third, two hybrid bases {\it mod4} and {\it mod5} are built. Each of them, being 
restricted by the maximal total number of the oscillator quanta $N^{max}_{tot}$  
involves the complete set of the 
NCSM WFs limited by inequality $N^{max}_{pol}\leq N^{max}_{tot}-2$. The sole 
additional cluster component with $N_{tot}=N^{max}_{tot}$ corresponding to {\it 
mod2} version is incorporated in {\it mod4}.
 For {\it mod5} version all cluster components with $N_{tot}=N^{max}_{tot}$ 
corresponding to {\it mod3} are incorporated. Evidently both {\it mod4} and {\it mod5}  
bases are incomplete in the space of the WFs with $N_{tot}=N^{max}_{tot}$.

 For each NN-potential the parameter $\hbar\omega$ demonstrating the best convergence of the
energy value of $^8$Be nucleus in NCSM calculations is chosen. Internal cluster WFs contained in the cluster 
terms are calculated in the framework of NCSM using the same NN-potential and the same 
$\hbar\omega$ parameter as the ones used for NCSM terms of eight-nucleon WFs.

\begin{figure}
\begin{center}\includegraphics[scale=0.45]{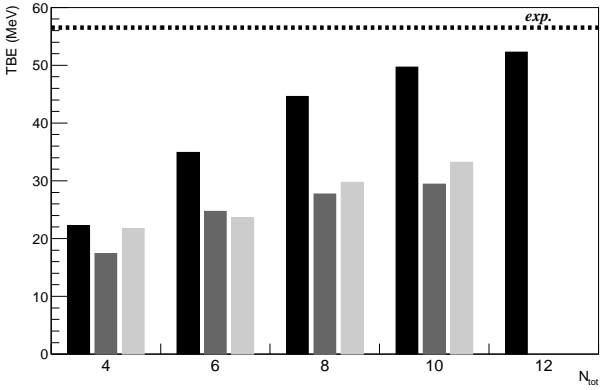}\end{center}
\caption{\label{fig1}TBE of $^8$Be nucleus calculated using JISP16 potential. 
Black columns -- {\it mod1}, dark-grey columns  -- {\it mod2}, pale-grey columns -- {\it mod3}. The
horizontal line shows the experimental TBE.}
\end{figure}

\begin{figure}
\begin{center}\includegraphics[scale=0.45]{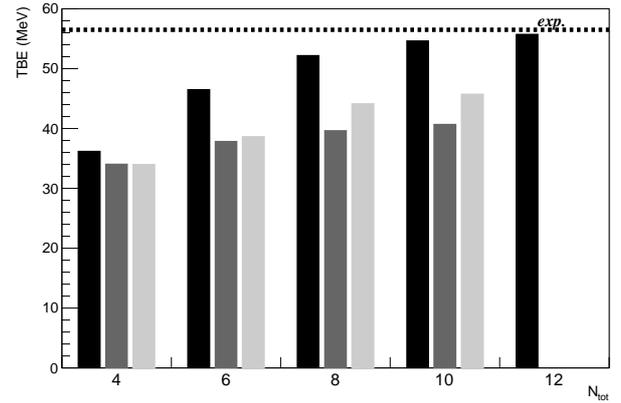}\end{center}
\caption{\label{fig2}TBE of $^8$Be calculated using Daejeon16 potential. Notations are
the same as in figure (\ref{fig1}).}
\end{figure}

The TBE of $^8$Be nucleus calculated in the framework of {\it mod1}, {\it mod2}, and
{\it mod3} models as a function of total number of the oscillator quanta $N_{tot}$ is
presented in figures \ref{fig1},\ref{fig2}. The former one have been computed using
the JISP16 potential with the oscillation parameter $\hbar\omega$ = 22.5 MeV, the latter one -- using
the Daejeon16 potential with the parameter $\hbar\omega$= 15.0 MeV.

In spite of significant difference between  two considered potentials, computations 
demonstrate very similar qualitative pattern. The cluster components corresponding to
lowest value of $N_{tot}$ contribute dominatingly compare to non-clustered ones with the
same truncation level both for one-channel {\it mod2} and three-channel {\it mod3}
models. However even in such a trivial basis the role of non-clustered components is not
negligible.  The extension of the basis makes the pattern drastic -- the relative
contribution of  components of such a type increases much rapidly than the clustered
ones. Thus the use of a basis consists from cluster WFs in the {\it ab initio}
approaches results in very large underestimation of the binding energy even though a
system under study is strongly clustered. This result is irrespective of whether the
lowest-configuration cluster model {\it mod2} or realistic cluster model {\it mod3} is
considered.

Properties similar to the just demonstrated were pointed out in lighter cluster systems. As is was shown in Ref. \cite{neff2} in the framework of FMD, thresholds of the cluster channels in seven-nucleon systems $^7$Li and $^7$Be turned out to be about 1 MeV  lower compare to the experimental ones in the case that "frozen" cluster configuration are considered. Including of intrisic basis states (polarization terms) remove this gap. Three-body system $\alpha$+n+n was considered inrecently published paper \cite{ncsmc}. TBE of this systems, computed in the cluster approach is  underestimated by 1 MeV in comparison with the results of accurate NCSMC calculation. It should be noted, however, that the effect detected in the current work for  $^8$Be nucleus is much greater. 

It makes sense to compare the quantitative results obtained by application of different
potentials. The Daejeon16 version being essentially more "soft" than JISP16 brings
about larger contribution of the cluster components to the TBE compared to JISP16. This
trend is confirmed in the case that the results of Ref. \cite{kv}, in which "supersoft"
version of  N3LO-based potential \cite{bfp} is exploited, to take into account.
Nevertheless the contribution of non-clustered components to $^8$Be TBE remains distinct
although it is noticeably smaller. A confirmation of the trend for a lighter cluster sistem may be found in just mentioned Ref.  \cite{ncsmc}. 

To perform more detail analysis of relationship between the contributions of cluster and
non-clustered components into the TBE we have carried out the calculation of its values
in {\it mod4} and {\it mod5} versions. This study allows one to compare the
contributions of these two types  of components at fixed values $N_{tot}$. To be brief
we present here the results of calculations in which Daejeon16 potential has been used.
The corresponding results obtained by use of JISP16 potential  are qualitatively 
 analogous.
 
The results are presented in Tab. \ref{tab-be-d}. The analysis of difference in the TBE
values between the ones computed in the mixed bases {\it mod4} or {\it mod5} with the
truncation level $N_{tot}$ and ones computed with the use of of NCSM {\it mod1}
characterized by the truncation level values $N_{tot}-2$ as well as $N_{tot}$ is the
most informative. In all cases extension of the NCSM basis restricted at the level
$N_{tot}-2$ by including the complete set of the NCSM WFs with $N_{tot}$ changes the
TBE value much stronger than the addition of  only the extra cluster components with the
same $N_{tot}$. For large values of $N_{tot}$ the cluster components increase the TBE
by few hundreds keV while involving the complete set of the components corresponding to
$N_{tot}$ adds several MeV. Thus the aggregate contribution of non-clustered components
to TBE in this area of $N_{tot}$ values is dominating. Moreover the trend is that their
relative contribution grows with the increasing of $N_{tot}$.

\begin{table}
\caption{TBE (MeV) of $^8$Be nucleus calculated using Daejeon16 potential and different bases with
$\hbar\omega$= 15.0 MeV.}\label{tab-be-d}
\begin{tabular}{cccccc}
\hline\hline\noalign{\smallskip} mod & N=4 & N=6 &N=8 &N=10 &N=12  \\
\hline\noalign{\smallskip}
mod1 &36.20&46.47&52.17&54.62&55.72 \\
mod4 &&39.00&47.16&52.34&\\
mod5 &&38.98&46.82&52.22& \\
\noalign{\smallskip}\hline\hline
\end{tabular}
\end{table}

The AoC, being one-channel or aggregate, serves as a direct measure of cluster content
of nuclear states. The computations of this values have been performed in the current
paper starting from the following points. The WF of initial nucleus $^8$Be ground state
was calculated in the framework of NCSM. The terms of the cluster channels (\ref{eq1})
are written in the form containing internal WFs of $\alpha$ clusters which in turn are
calculated in the basis limited by the condition $N^{max}=2$. The channels corresponding
to the ground states of both clusters comprising a nucleus are considered together with
the channels which contain one or both $\alpha$-particles in $0^+_2$ state. Higher
excitations of $^4$He nucleus have very large values of the decay width to be considered
as realistic clusters. In Tab. \ref{t-sf-be-d} we present the values of one-channel
(ground-ground) and aggregate two-channel (plus ground-excited), and all three-channel
(plus excited-excited) AoC.

\begin{table}
\caption{AoC in $^8$Be nucleus calculated using Daejeon16 potential and the {\it mod2} basis with
$\hbar\omega$= 15.0 MeV.}\label{t-sf-be-d}
\begin{tabular}{cccccc}
\hline\hline\noalign{\smallskip} mod & N=4 & N=6 &N=8 &N=10 &N=12  \\
\hline\noalign{\smallskip}
one-ch  &0.068&0.765&0.866&0.861&0.875 \\
two-ch &0.442&0.793&0.868&0.868&\\
three-ch &0.992&0.864&0.879&0.873&\\
\noalign{\smallskip}\hline\hline
\end{tabular}
\end{table}

The table demonstrates a rapid convergency of the AoC values. The saturation is achieved at the level Ntot=8 with reasonable precision. The small value of one-channel AoC for Ntot=4 is an artefact of the chosen basis. Indeed, the quantum number $n$=0 characterising the relative motion wave function is contained for this choice of  $N^{max}_{tot}$.
The numerical results of the
one-channel {\it ab initio} calculations  of SFs with realistic WFs of the clusters as
well as the results of Ref. \cite {kv} favour the view that the system under study is
strongly clustered. All these statements based on the results of {\it ab initio}
calculations provide a support for a variety of microscopic models of "nuclear cluster
geometry", based on clustered probe WFs and the dynamics describing by effective
nucleon-nucleon Hamiltonians, such as AMD \cite{keh}, THSR-approach \cite{thsr}, etc.
(this "geometry" manifests itself in a specific density distribution of a clustered
system in the internal coordinate system).  Indeed, the clustered components constitutes
the major portion of the realistic WF. At the same time another implication of this
analysis is that the use of realistic NN-potentials would result in significant
disagreement between TBE values computed in the clustered models and the experimental
ones. Effective NN-potentials turn out to be necessary to compensate this disagreement.
Besides that it is preferable realistic internal cluster WFs to be used in such
approaches.

The results  of the current study contained some other unexpected points.

Let us consider, first of all, relationship between the values of AoC in one-channel and
multi-channel cases which are illustrated by Tab. \ref{t-sf-be-d}. In the region of
saturation of the results the aggregate AoC value is ever so slightly greater than the
one-channel AoC. Consequently the extremely strong non-orthogonality of the WFs of the
discussed channels takes place and turns out to be the reason of that. Thus the
procedure of extension of cluster-channel basis is slow in affecting the convergency of
both eigenvalues and eigenvectors of {\it ab initio} Hamiltonians in spite of the
completeness of the basis.

Another point have been detected in the simultaneous analysis of TBE and AoC. We stress, that taking 
into account that the terms of the cluster basis  are non-orthogonal to NCSM wave 
functions, non-clustered  components are defined as all components of the basis 
obtained by the orthonormalization procedure (\ref{eq8}) besides the considered cluster ones. 
Denoting aggregate AoC as $B$ one can define the statistical weight of non-clustered components
as $\bar B=1-B$. Introducing the ratio $\delta E=(E^{mod1}-E^{mod2(3)})/E^{mod1}$ it is
possible to define the measure of contribution of non-clustered components into the TBE.

A comparison between these values deduced from the respective tables demonstrate that
$\delta E>\bar B$ for larger values of $N_{tot}$. Thus the contribution of non-clustered
components to the TBE related to their statistical weight is greater than the
corresponding relative contribution of the cluster ones and this trend is the most
expressive at larger values of $N_{tot}$.

Contrastingly lighter system $^6$He manifests other properties. According to Ref. \cite{ncsmc} AoC of the ground-state $\alpha$-particle channels being approximately equal to the ones presented in Tab. \ref{t-sf-be-d} contributes, as it has been noted above,  only 1 MeV to TBE of the system.

Summing all the presented up a brief review of the basic points should be made.

\noindent 1. A formalism  convenient to construct a basis that takes into consideration
the cluster properties of nuclear systems in {\it ab initio} calculations of their
characteristics is built. The keystone of the formalism is the technique of the cluster
coefficients. The developed approach is universal. Various versions of the basis can be
applied to conform to the cluster properties of systems under study.

\noindent 2. An additive and normalized to unity measure of cluster content -- amount of
clustering \cite{lov} -- is extended to be used for studies of the multi-channel
problem.

\noindent 3. The total binding energies of $^8$Be nucleus as well as the values of one-
and multi-channel amount of clustering are computed. As the result of the combined 
analysis of these values a fuller picture of a realistic clustered state as a pure
cluster configuration immersed into a "sea" of diffuse non-clustered components is
beginning to emerge. The statistical weight of these components is shown to be moderate
but their contribution to the TBE is great.

The work was supported by Russian Science Foundation (RSF), grant No. 16-12-10048.
Authors are grateful to A. M. Shirokov for fruitful discussions.

\bibliographystyle{model3-num-names}

\bibliography{sample}

\end{document}